\documentclass[twocolumn,showpacs,preprintnumbers,amsmath,amssymb]{revtex4}

\usepackage[latin1]{inputenc}
\usepackage{graphicx}

\begin{document}

\title{Consciousness, brains and the replica problem}

\author{Ricard V. Solé}
\email{ricard.sole@upf.edu}

\affiliation{$^1$ ICREA-Complex Systems Lab, Universitat Pompeu Fabra. 
Parc de Recerca Biomedica de Barcelona.  
Dr Aiguader 88, 08003 Barcelona, Spain}

\begin{abstract}
Although the conscious state is considered an emergent property of the underlying 
brain activity and thus somehow resides on brain hardware, 
there is a non-univocal mapping between both. Given a neural hardware, 
multiple conscious patterns are consisten with it. Here we show, by means 
of a simple {\em gedankenexperiment} that this has an importan logic 
consequence: any scenario involving the transient shutdown of brain activity 
leads to the irreversible death of the conscious experience. In a fundamental way, 
unless the continuous stream of consciousness is guaranteed, the previous 
self vanishes and is replaced by a new one. 
\end{abstract}


\maketitle

\section{Introduction}

The problem of consciousness has become a hot topic of scientific enquiry over the 
last two decades (Searle, 2000, Crick and Koch, 1995, 2003). But in spite of this increasing 
attention from the neurosciences, old questions remain open and the phenomenon itself 
differs from other biological phenomena in that it is a subjective, first-person ontology (Searle, 2000). 
Such special status generates a number of nontrivial questions, some of them right in the 
boundaries between science and philosophy. Most neuroscientists, with few exceptions, would agree 
(even with different perspectives) that consciousness 
is a self-organized, emergent property of brain activity and neuronal
 wiring, although the nature and organization of the brain-mind mapping 
is largely unknown (Locke, 1995; Dennett, 1991; Hesslow, 1994; Svenson, 1994; 
von Wright, 1994; Crick and Koch, 2003). Multiple 
questions emerge from the previous scenarios, including the nature of the 
new consciousness emerging after recovery from long-term cryogenization or technological 
replacement (Moravec, 1988; Egan, 1994; Minsky 1994). Similar problems arise in different contexts, such as 
teleportation (Penrose, 1989). How can a transient shutdown of brain activity affect 
the conscious experience? All the previous situations inhabit the realm of speculation 
and might never be achieved. The potential implications are mostly a matter of philosophical speculation. 
There is, however, an experimentally feasible scenario 
where no such speculation is at work. 

Recently, advances in suspended 
animation suggest the possibility of preserving human life in a reversible 
state where completely halted or deeply slowed cellular activity would be possible (Alam et al., 2005). 
Such state has been obtained experimentally using different organisms 
(Nystul and Roth, 2004; Blackston et al., 2005) and nothing prevents to reach similar results 
using humans. In fact, evidence from accidental, long-term suspended animation 
is available from a number of case studies. In these cases, humans experiencing severe 
hpothermia over several hours and showing lack of any vital sign (no pulse nor 
brain activity) were able to recover without any long-term complications. Ongoing 
research on using profound hypothermia, together with appropriate organ preservation 
fluids confirm that such reversible states can be induced in a repeatable manner 
(Alam et al., 2005). The method, used in swine animal models, 
results in clinical brain death, but none of the 
surviving hypothermic animals displayed detectable neurological deficits or cognitive 
impairment.

How can a shutdown of brain activity alter the nature of the self-conscious experience? 
In principle, you might think that your consciousness is temporally stopped, just 
to be back afterwards. In other words, you and your consciousness weak up altogether. 
Is that really the case? To put the question in a more specific form, we 
consider a mental (Gedanken) experiment, which we will call {\em the replica problem}.
Below we show, using a logic argument, that something much more fundamental is at work 
when considering scenarios involving consciousness and its relation to hardware. 
Together with brain death (no matter if permanent of transient) 
the death of subjective consciousness needs to be considered.

\begin{figure*}
\includegraphics[width=12cm]{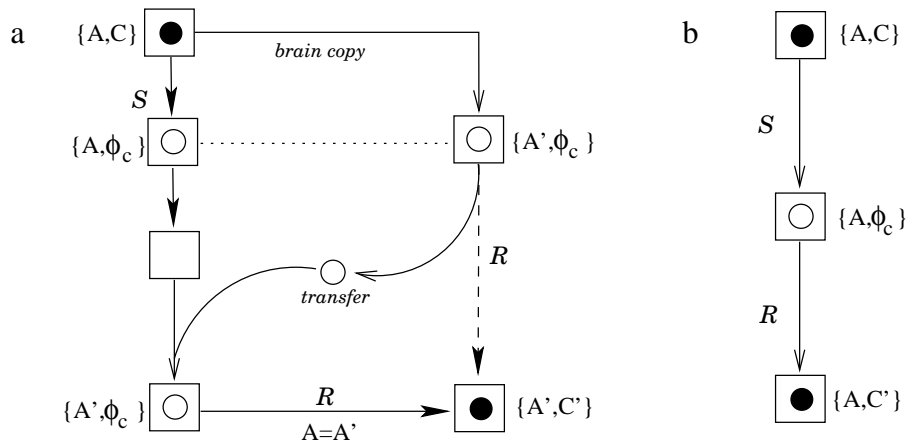}
\vspace{0.5 cm}
\caption{(a) The extended replica problem, as defined in the text. Here we start with 
an individual defined as the brain-mind pair $\{A,C\}$. A copy of the 
brain hardware is made, with no activity and thus no consciousness, here indicated 
as $\{ A', \phi_c \}$. Since the new physical hardware is an exact copy, no 
experiment would be able to distinguish between $A$ and $A'$. 
If activated (dashed line, lower right) the copied system would obviously display a separated conscious 
experience, here indicated as $C'$. If the $A$'s brain is extracted and replaced 
by $A'$, we would have exactly the same hardware (so effectively $A=A'$) and 
no difference would be measurable. However, once activated again, it would 
not exhibit the initial subjective conscious experience, but a different one. The previous 
experiment is equivalent to the situation shown in (b) where we simply shut down brain 
activity and afterwards reverse the unconscious state into a conscious one.} 
\end{figure*}

\section{The replica problem}

The following experiment is an imaginary one, not expected to be ever possible. It is thus 
a Gedankenexperiment, used as a logic argument to show the unexpected consequences 
of the one-to-many brain-mind mapping. It is important to stress that this is a thought 
experiment and is thus not expected to be possible. In this context, we are aware that quantum mechanics 
forbids the realization of the ideal experiment to be described below (Scarani et al., 2005) 
but that is not relevant 
to our discussion, particularly because quantum effects should not be expected to 
have a real relevance in large scale neural dynamics. However, although the special case considered 
here would require a high-level technology not available today, some equivalent 
scenarios (such as the induction of profound hypothermia discussed above) 
are likely to be soon applied to human beings.   

 Let us take a given individual brain $A$, experiencing a given (self-)conscious activity. 
We can indicate that the conscious experience $C$ is somehow generated by 
this brain $A$ using a mapping:
\begin{equation}
A  \longrightarrow C
\end{equation}
Where $C$ must be interpreted as an emergent property of brain activity and involves 
both subjectivity and self-awareness. Let us now imagine that thanks to a very 
advanced technology a full copy of 
$A$ can be obtained instantaneously at $t = t_0$. Considering instantaneous formation is not 
strictly necessary, but makes the argument simpler, since it liberates 
us from considering the further divergence of the two replicated systems. 
Let us call this new brain hardware $A'$. This replica, if active, would generate 
a different conscious experience, which we indicate as $C'$. Clearly we have now:
\begin{equation}
A' \longrightarrow C'
\end{equation}
the important point here is that, 
although {\em exactly the same hardware} is being used, we have $C \ne C'$ (different 
subjective conscious experiences). This is true in 
spite that no single experiment made by some external observer would be able (at $t = t_0$) 
to distinguish between $A$ and $A'$. The existence of a replica of $A$ generates 
a somewhat strange situation, since clearly indicates that brain 
activity does not univocally define consciousness. This is what we 
name {\em the replica problem}. This problem has been explored by a number 
of authors (see http://www.benbest.com/philo/doubles.html) and is our starting point. 

Let us now consider $A$, with an associated conscious experience $C$. The 
brain-mind pair $\{A,C\}$ thus fully defines the individual. Let us assume that brain activity is stopped 
through some process $S$. If no brain activity is present, no conscious experience exists. The individual's 
brain is dead, and will be indicated as $\phi_c$, meaning 'no consciousness' (here the symbol $\phi_c$ 
indicates lack of consciousness, without explicit reference to a given $C$). 
Now let us imagine that the brain is reactivated through some other process $R$. The standard view 
considers the following causal set of events:
\begin{equation}
\{A,C\} \buildrel S \over  \longrightarrow \{A,\phi_c \} \buildrel R \over  \longrightarrow \{A,C \}
\end{equation}
This logical chain of events corresponds to a common reasoning: my brain is freezed and stops 
working, but once a reverse process is used, brain activity returns and I wake up.  
Is that a correct answer? Which consciousness is 
experienced: the previous one ($C$) or a new one ($C'$)? As shown below, a 
new consciousness is effectively at work, i. e. the correct sequence is in fact: 
\begin{equation}
\{A,C\} \buildrel S \over  \longrightarrow \{A,\phi_c \} \buildrel R \over  \longrightarrow \{A,C' \}
\end{equation}
and thus, in terms of consciousness, we never ``wake up''. 
The reason is that the hardware does not univocally 
define the conscious experience, and thus there is no reason why the conscious 
activity emerging after recovering the stopped brain would be the same. 
However, you might argue that it is the same brain what is at work, and thus 
cannot be properly related with the replica problem, where two identical, 
but different brains are being used.

An additional experiment allows to better understand the implications of the 
replica problem. This {\em extended replica problem} can be used to see clearly why 
the new conscious experience is necessarily a different one. The basic steps to 
be described below are summarized in figure 1.

\section{The extended replica problem}

We now describe a special mental experiment involving the formation of a 
replica. In figure 1, individuals involving an active (and thus conscious) brain 
are indicated as framed black circles. If brain activity is stopped, the non-conscious 
state is indicated as an empty circle. If no brain is present, an empty box is shown. 

Let us assume that we start with 
$\{A,C\}$ and we make a material (but not active) copy $A'$ of the initial brain. We have a 
new brain-mind system $\{A',\phi_c \}$ with no consciouss activity ($\phi_c$) and physically 
separated from the initial one (see upper part of figure 1a). If activated, 
$A'$s brain would generate its own subjective conscious state, i. e. 
\begin{equation}
\{ A', \neg C \} \buildrel R \over  \longrightarrow \{ A', C' \}
\end{equation}
with $C'$ obviously different from $C$ (lower right, fig. 1a). 
Now we shut down the activity of $A$ i.e. 
\begin{equation}
\{ A, C \} \buildrel S \over  \longrightarrow \{ A, \phi_c \}
\end{equation}
And now let us replace $A$ by $A'$, i. e. 
\begin{equation}
\{ A, \phi_c \} \rightarrow \{ A', \phi_c \}
\end{equation}
Since the two brains are physically identical, no measurement would be able to detect 
any difference between the previous and the new hardware, and thus we have the equivalence: 
\begin{equation}
\{ A, \phi_c \} \equiv \{ A', \phi_c \}
\end{equation}
The logic implication is that they can be exchanged by each other (and any other exact copy) and 
would not be distinguished. But it is know obvious that the implanted brain, though identical, is not 
going to maintain the subjective conscious experience that we had at the beginning: 
it was a copy and following the previous implications we would have
\begin{equation}
\{ A, \phi_c \} \buildrel R \over  \longrightarrow \{ A, C' \}
\end{equation} 
The sequence of events described above is logically equivalent to starting from 
$\{A,C\}$, stopping $A$ from being active and restoring its function (ending up in 
$\{ A, C' \}$, as indicated in figure 1b. 
This completes our argument. To summarize: any process that either stops brain activity 
(and thus leaves us with a ``just hardware'' individual) or replaces a given 
brain structure by a completely new one (after stopping consciousness in its previous 
physical support) leads to a state of ``dead consciousness''. As a consequence of the non-unique mapping 
between brain structure and conscioussness, death of a given conscious experience 
will be irreversible.

\section{Discussion}

In this paper we have seen how the one-to-many mapping between brain and mind 
implies that any scenario involving transient brain death leads to the death 
of consciousness, as defined by a subjective, first-person ontology. 
The subjective nature of the self makes brain transfer and 
teleportation non-viable in terms of a reliable way of transfering the 
self to the new individual. These are, however, science fiction scenarios. However, 
as shown above, the same situation must be applied to surgery involving profound hypothermia: 
you (meaning your self) would never truly wake up once the normal brain 
function is recovered again. Someone else will, 
with exactly the same external features and memories as you, but experiencing a 
different consciousness. Under this view, no true immortality (the immortal 
nature of your self) is possible. 

Although future technology might allow building a copy of our brains and 
make our memories and feelings survive, something will be inevitably lost. 
The argument provided here suggests that the ``self'' persists (it is {\em alive}) provided that the 
stream of consciousness flows in a continuous manner and is never interrupted. 
If it is, death of the self occurs in a non-reversible manner. This seems 
to provide an interesting twist to the mind-body problem.

Although the argument presented here is a logical one, further extensions of this 
study would involve brain states not necessarily associated to a complete 
lack of activity. More quantitative analyses could be made, involving different 
features of consciousness (Seth et al., 2006) and the possible localization of the 
conscious self-representation (Lou et al., 2004 and references therein). In this context, 
further questions arise: What are the minimum requirements in terms of brain activity able 
to sustain a conscious pattern? Are there partial changes inducing a loss of 
self-awareness related to our previous discussion?

\begin{acknowledgments}

The author would like to thank the members of the Complex Systems Lab 
for useful discussions. Special thanks to Bernat Corominas-Murtra 
for comments on the manuscript. I also thank the editor and referees of 
{\em Minds and Machines} who kindly rejected this paper with no rational explanation. 

\end{acknowledgments}


\vspace{0.5 cm}

{\bf References}

\vspace{0.5 cm}

\begin{enumerate}

\item
Alam, H. B. et al. 2005. Profound Hypothermia Protects Neurons and Astrocytes, and Preserves 
Cognitive Functions in a Swine Model of Lethal Hemorrhage. J. Surg. Res. 126, 172-181.\\

Blackstone, E., Morrison, M. and Roth, M. B. 2005. Hydrogen sulfide induces 
a suspended animation-like state in mice. Science 308, 518.\\

Brooks, R. A. 2002. {\it Flesh and machines: How Robots Will Chage Us}. Pantheon Books, New York.\\

Crick, F. C. and Koch, C. 1995. Why neuroscience may be able to explain consciousness. Sci. Am. 273, 84-85.\\

Crick, F. C. and Koch, C. 2003. A framework for consciousness. Nature Neurosci. 6, 119-126.

Dennett, D. C. 1991. {\it Consciousness explained}. Little, Brown and Company, Boston. \\ 

Egan, G. 1994.  {\it Permutation City}. Milennium , London.\\

Hesslow, G. 1994. Will neuroscience explain consciousness? J. Theor. Biol. 171, 29-39.\\

Locke, J. 1995. {\em An Essay Concerning Human Understanding}. Prometheus Books, New York.\\

Lou, H. C. 2004. Parietal cortex and representation of the mental self. Proc. Natl. Acad. Sci. USA 101, 6827-6832.\\

Minsky, M. 1994. Will robots inherit Earth? Sci. Am. 271, 108-13 \\

Moravec, H. 1988. {\em Mind children. The future of robot and human intelligence}. 
Harvard U. Press, Harvard.\\

Nystul, T. and Roth, M. B. 2005. Carbon monoxide-induced suspended animation protects 
against hypoxic damage in Caenorhabditis elegans. Proc. Natl. Acad. Sci. USA 101, 9133-9136.\\

Penrose, R. 1989. {\em The emperor's new mind}. Vintage Books, London.\\

Safar P, Tisherman SA, Behringer W, Capone A, Prueckner S, Radovsky A, Stezoski WS, Woods RJ. (2000) 
Suspended animation for delayed resuscitation from prolonged cardiac 
arrest that is unresuscitable by standard cardiopulmonary-cerebral resuscitation.
Crit. Care Med. 28 (Suppl), N214-218.\\

Scarani, V., Iblisdir, S. and Gisin, N. 2005. Quantum cloning. Rev. Mod. Phys. 77, 1225-1256.\\

Seth, A. K., Izhikecih, E., Reeke, G. N. and Edelman, G. M. (2006) Theories and measures of 
consciousness: an extended framework. Proc. Natl. Acad. Sci. USA 103, 10799-10804.\\

Svensson, G. (1994) Reflections on the problem of indentifying mind and brain. 
J. Theor. Biol. 171, 93-100.\\

Tisherman, S. A. (2004) Suspended animation for resuscitation from exsanguinating hemorrhage. 
Crit. Care Med. 32(2 Suppl), S46-50.\\

von Wright, G. H. (1994) On mind and matter. J. Theor. Biol. 171, 101-110.

\end{enumerate}


\end{document}